\documentclass[12pt]{article}

\usepackage[square,sort,comma,numbers]{natbib}
\usepackage{graphicx}
\usepackage{amssymb}
\usepackage{amsfonts}
\usepackage{amsmath}

\usepackage{times}

\newenvironment{sciabstract}{%
\begin{quote} \bf}
{\end{quote}}

\newcounter{lastnote}

\newcommand{\TG}[1][]{T_{\mathrm{G}#1}}
\newcommand{\TB}[1][]{T_{\mathrm{B}#1}}
\newcommand{\SG}[1][]{S_{\mathrm{G}#1}}
\newcommand{\SB}[1][]{S_{\mathrm{B}#1}}
\newcommand{\pG}[1][]{p_{\mathrm{G}#1}}
\newcommand{\pB}[1][]{p_{\mathrm{B}#1}}

\newcommand{\TCold}{T^{\mathrm{C}}}
\newcommand{\THot}{T^{\mathrm{H}}}

\newcommand{\kB}{k_{\mathrm{B}}}
\newcommand{\Tr}{\mathrm{Tr}}

\title{Inconsistent thermostatistics and\\  negative absolute temperatures} 

\author
{J\"orn Dunkel$^{1\ast}$ 
and
Stefan Hilbert$^{2}$\\
\\
\normalsize{$^{1}$Department of Applied Mathematics and Theoretical Physics, University of Cambridge,}\\
\normalsize{Wilberforce Road, Cambridge CB3 0WA, UK}\\
\normalsize{$^{2}$Max Planck Institute for Astrophysics, Karl-Schwarzschild-Str. 1, 
85748 Garching, Germany}\\
\\
\normalsize{$^\ast$To whom correspondence should be addressed; E-mail:  jd548@cam.ac.uk}
}

\date{}

\begin{document} 




\maketitle


\begin{sciabstract} 
A considerable body of experimental and theoretical work claims the existence of negative absolute temperatures in spin systems and ultra-cold quantum gases. Here, we clarify 
that such findings can be attributed to the use of a popular yet inconsistent entropy definition,
which violates fundamental thermodynamic relations and fails to produce sensible results for simple analytically tractable classical and quantum systems. Within a mathematically consistent thermodynamic formalism, based on an entropy concept originally derived by Gibbs, absolute temperature remains positive even for systems with  bounded spectrum. We address spurious arguments against the Gibbs formalism and comment briefly on heat engines with efficiencies greater than one.
\end{sciabstract}

\section{Introduction}

 The notion of \lq negative absolute temperature\rq~appears to have been first introduced by Purcell and Pound~\citep{1951PurcellPound} in 1951 to describe population inversion in nuclear spin systems. A few years later, in 1956, their ideas were more broadly formalized by Ramsay~\citep{1956Ramsay}, who discusses several ramifications of negative temperature systems\footnote{Somewhat paradoxically perhaps, systems that are claimed to possess a negative absolute  temperature are known to be \lq hotter\rq~than their positive-temperature counterparts~\citep{1994Hakonen_Science}, and one is tempted to wonder why within a consistent thermodynamic description this fact should not be directly evident from the temperature itself.}, most notably the hypothetical possibility to create Carnot machines with efficiencies larger than one~\citep{1977Landsberg_JPhysA,2010Rapp} -- which, if achievable in a meaningful manner, would solve all  future energy problems. More recently, the experimental realization of an ultra-cold bosonic quantum gas with bounded spectrum~\citep{2013Braun} has attracted considerable attention as another apparent example of a system with negative absolute temperature~\citep{2013Carr}, encouraging speculation~\citep{2013Braun} that negative temperature states could be of interest as Dark Energy candidates in cosmology.  
 
 \par
Our discussion below will show that  negative absolute temperatures, as reported for  spin systems and quantum gases,  arise from the application of a popular yet inconsistent microcanonical entropy definition,  usually attributed to Boltzmann\footnote{It might be unjust to direct any form of criticism at Boltzmann here, as it is unclear whether or not he deemed this particular entropy definition applicable  in this context (see historical remarks on pp. 181 in Sommerfeld's book~\cite{Sommerfeld}).}.  Restricting our arguments exclusively to mathematically rigorous results~\citep{Khinchin} and exactly solvable examples,  we will demonstrate that the commonly used Boltzmann entropy is generally incompatible with the most basic thermodynamic relations, fails to give sensible results for simple analytically tractable quantum systems, and violates equipartition in the classical limit.

\par
These as well as several other deficiencies can be cured by adopting a self-consistent entropy concept that was derived  by J. W. Gibbs more than 100 years ago~\citep{Gibbs}, but seems to have been mostly forgotten ever since. Unlike the Boltzmann entropy, the Gibbs entropy produces intuitively reasonable predictions for heat capacities and other thermodynamic observables in all exactly computable test cases known to us, and it yields a non-negative absolute temperature even for quantum systems with  bounded spectrum.

\par
Despite its conceptual advantages, the Gibbs formalism is often met with skepticism that appears to be rooted in habitual preference of the Boltzmann entropy rather than objective evaluation of factual evidence. We therefore complement our more technical considerations (Sec.~\ref{s:MCE} and \ref{s:spin_example}) by addressing a number of frequently encountered spurious arguments against the Gibbs entropy (Sec.~\ref{s:myths}), and we also comment briefly on misconceptions about the possibility of heat engines with efficiencies  greater than one (Sec.~\ref{s:carnot}).

\par
To set the stage for the subsequent discussion, it is useful to review a standard argument in favor of  negative absolute temperatures, which seems rather plausible at first sight~\citep{2013Carr}:  Assume, a suitably designed many-particle quantum system with bounded spectrum~\citep{1951PurcellPound,2013Braun} and non-monotonous density of states (see Fig.~\ref{fig:thermodynamic_functions_for_weakly_coupled_bosonic_oscillators}) can be driven to a \emph{stable} state of population inversion, so that the majority of particles occupy high-energy one-particle levels. The one-particle energy distribution of such a system will be an increasing function of the one-particle energy $\epsilon$. In order to fit \citep{2013Braun} such a distribution with a Boltzmann factor\footnote{Or Bose or Fermi-functions, depending on the details of the system.}~$\propto \exp(-\beta \epsilon)$, one must have a negative $\beta$ and, hence, a negative Boltzmann temperature $\TB=(\kB\beta)^{-1}<0$. Whilst this reasoning may  indeed  appear straightforward, some reservation is in order as it is uncertain
\begin{itemize}
\item[(i)]  whether the one-particle energy distribution of a system with bounded spectrum is a simple exponential, or Bose, or Fermi function over the full spectral energy range, and, even if it were, 
\item[(ii)]  that the fit-parameter $\TB$  coincides with the absolute temperature of the many-particle systems. 
\end{itemize}
To appreciate these concerns, it is important to recognize that population-inverted states are generally thermodynamically unstable when coupled to a (non-population-inverted) environment. This means that such systems must be prepared in isolation and, hence, their thermodynamic description has to be based on the microcanonical ensemble.

\par
The arguments presented below clarify that the one-particle Boltzmann temperature $\TB$ is, in general, not the absolute thermodynamic (i.e., microcanonical) temperature $T$, unless one is willing to abandon the defining differential relations of thermodynamics and thermostatistics. Moreover, by presenting a mathematically exact relation between $\TB$ and $T$, we verify that the absolute temperature $T$ remains positive even in the case of a bounded spectrum with $\TB<0$.

\section{Microcanonical entropy revisited}
\label{s:MCE}

When interpreting thermodynamic data of exotic many-body states~\citep{2013Braun},  one of the first questions that needs to be addressed is the choice of the appropriate thermostatistical ensemble~\citep{2009Campisi_PRL,2010Campisi}. Equivalence of the microcanonical (MC) and other canonical ensembles cannot -- in fact, must not -- be taken for granted for systems that are characterized by a non-monotonic~\citep{1956Ramsay,2010Rapp,2013Braun} density of states (DoS) or that can undergo phase-transitions due to attractive interactions~\citep{2006DuHi} -- gravity being a prominent example~\citep{2002Gross_Full}. For instance, for an ultracold quantum gas~\citep{2013Braun}  that has been isolated from the environment to suppress decoherence, both particle number and energy are in good approximation conserved. Therefore, barring other  physical or topological  constraints,  any {\it ab initio} thermostatistical treatment should start from the MC ensemble.

\par
In this section, we will first prove that the Gibbs entropy provides a consistent thermostatistical model for the MC density operator, which also implies that the popular Boltzmann entropy is not the thermodynamical entropy of the MC ensemble. We then illustrate the deficiencies of the Boltzmann entropy with a number of explicit test examples.  Last but not least, we still explain why the one-particle distribution measured by Braun {\it et~al.}~\citep{2013Braun} features the effective Boltzmann temperature $\TB$ and not the thermodynamic absolute temperature $T$, which is determined by the Gibbs entropy. An example relevant to the experiments of Purcell and Pound~\citep{1951PurcellPound} and Braun {\it et~al.}~\citep{2013Braun} is discussed in  Sec.~ \ref{s:spin_example}.

\subsection{Entropy and temperature definitions}
\label{s:MCE_S_and_T}

To make the discussion more specific, let us consider a (quantum or classical) system with microscopic variables~$\xi$ governed by the Hamiltonian $H=H(\xi;V,A)$, where $V$ denotes volume and  $A=(A_1,\ldots)$ summarizes other external  parameters. Assuming that the dynamics conserves the energy, $E=H$, all thermostatistical properties are contained in the MC density operator\footnote{As usual, we assume that,  in the case of quantum systems, Eq.~$\eqref{e:MCE_rho}$ has a well-defined operator interpretation.} 
\begin{equation}
\label{e:MCE_rho}
\rho(\xi;E,V,A)= \frac{\delta(E-H)}{\omega},
\end{equation}
which is normalized by the DoS
\begin{equation}
\omega(E,V,A)=
\Tr [\delta(E-H)].
\end{equation}
For classical systems, the trace simply becomes a phase-space integral over $\xi$. For brevity, we denote averages of some quantity $F$ with respect to the MC density operator $\rho$ by $\langle F \rangle \equiv \Tr[ F\rho]$.
\par
We also define the integrated DoS\footnote{Intuitively, for a quantum system with spectrum $\{E_n\}$, the quantity $\Omega(E_n,V,A)$ counts the number of eigenstates with energy less or equal to $E_n$.} 
\begin{equation}
\Omega(E,V,A)=
\Tr [\Theta(E-H)],
\end{equation} 
which is related to the DoS $\omega$ by differentiation with respect to energy,
\begin{equation}
\omega=\frac{\partial \Omega}{\partial E}\equiv \Omega'.
\end{equation}

\par
Given the MCE density operator~\eqref{e:MCE_rho}, one can find two competing definitions for the MC entropy in the literature~\citep{Khinchin,Gibbs,2006DuHi,Becker,Huang,2007Campisi}
\begin{align}
\label{e:S_B}
\SB(E,V,A) &= \kB \ln [\epsilon \omega(E)],
\\
\SG (E,V,A)&= \kB \ln [\Omega(E)],
\label{e:S_G}
\end{align} 
where $\epsilon$ is a constant with dimensions of energy, required to make the argument of the logarithm dimensionless\footnote{Apart from other more severe shortcomings of $\SB$, it is aesthetically displeasing that its definition requires the~\textit{ad hoc} introduction of some undetermined constant.}. The first proposal, $\SB$, usually referred to as Boltzmann entropy, is advocated by the majority of modern textbooks~\citep{Huang} and used by most authors nowadays.  The second candidate $\SG$ is often  attributed to P. Hertz\footnote{Hertz proved in 1910 that $\SG$ is an adiabatic invariant~\citep{1910Hertz}. His work was highly commended by Planck~\citep{2008Hoffmann} and Einstein, who closes his comment~\citep{1911Einstein} on Hertz's work  with the famous statement that he himself would not have published certain papers, had he been aware of Gibbs' comprehensive treatise~\citep{Gibbs}.}~\citep{1910Hertz} but was in fact already derived by J. W. Gibbs  in 1902 in his discussion of thermodynamic analogies \citep[][Chapter XIV]{Gibbs}. For this reason,  we shall refer to $\SG$ as Gibbs entropy in the remainder. Denoting partial derivatives  of $\Omega$ and $\omega$ with respect to $E$ by a prime, the associated temperatures are given by
\begin{align}
\label{e:T_B}
\TB(E,V,A) &= 
 \left(\frac{\partial \SB}{\partial E}\right)^{-1}  
 = \frac{1}{\kB} \frac{\omega}{\omega'}=\frac{1}{\kB} \frac{\Omega'}{\Omega''},
\\
\TG (E,V,A)&=  
\left(\frac{\partial \SG}{\partial E}\right)^{-1}  
 = \frac{1}{\kB} \frac{\Omega}{\Omega'} = \frac{1}{\kB} \frac{\Omega}{\omega}.
\label{e:T_G}
\end{align} 
Note that $\TB$ becomes negative if $\omega'<0$, that is, if the DoS is non-monotic (Fig.~\ref{fig:thermodynamic_functions_for_weakly_coupled_bosonic_oscillators}), whereas $\TG$ is always non-negative, since $\Omega$ is a monotonic function of $E$.

\subsection{Consistency requirements}
\label{s:MCE_consistency}

The fundamental thermodynamic potential of the MCE is the entropy $S$, from which secondary thermodynamic observables, such as temperature $T$ or pressure $p$, are obtained
by differentiation with respect to the natural variables $E$, $V$, and $A$, i.e., the control parameters of the ensemble. The fundamental relation between entropy, control paramerers, and secondary thermodynamic variables can be expressed by
\begin{equation}
\begin{split}
\label{e:fundamental}
dS &=
\left(\frac{\partial S}{\partial E}\right) dE+\left(\frac{\partial S}{\partial V}\right) dV+\sum_{i} \left(\frac{\partial S}{\partial A_i}\right)dA_i,
\\&\equiv 
 \frac{1}{T} dE+\frac{p}{T} dV+\sum_{i} \frac{a_i}{T} dA_i.
\end{split}
\end{equation}
To form a \emph{consistent} thermostatistical model $(\rho,S)$, the entropy $S$ \emph{must} be defined such that   
the fundamental differential relation~\eqref{e:fundamental} is fulfilled\footnote{Our discussion is based on the premise that any acceptable thermostatistical model, corresponding to a pair $(\rho,S)$ where $\rho$ is a probability density and $S$ an entropy potential,  must satisfy Eq.~\eqref{e:fundamental}. If one is willing to abandon this requirement, then any relation to thermodynamics is lost.}.

\par
Equation~\eqref{e:fundamental} imposes stringent constraints on possible entropy candidates. For example, for an adiabatic (i.e., isentropic) volume change with $dS=0$ and other parameters fixed $(dA_i=0)$, one finds the consistency condition
\begin{equation}
\label{e:consistency_V}
p= T \left(\frac{\partial S}{\partial V}\right) =-\left(\frac{\partial E}{\partial V}\right).
\end{equation}
More generally, for any parameter $A_\mu\in\{V,A_i\}$ of the Hamiltonian $H$, one must have
\begin{equation}
\label{e:consistency_general}
a_\mu\equiv
-\left\langle  \frac{\partial H }{\partial A_\mu}\right\rangle
\equiv
-\Tr \biggl[\left(\frac{\partial H }{\partial A_\mu}\right) \rho\biggr]
\stackrel{!}{=}
T \left(\frac{\partial S}{\partial A_\mu}\right),
\end{equation}
where $T \equiv \left(\partial S/\partial E\right)^{-1}$.
These conditions not only ensure that the thermodynamic potential $S$ fulfills the fundamental differential relation~\eqref{e:fundamental}. They can also be used to separate consistent entropy definitions from inconsistent ones.
\par
Using simply the properties of the MC density operator, one derives from the above requirements that the MC entropy $S$ equals the Gibbs entropy $\SG$:
\begin{equation}
\begin{split}
\label{e:gibbs_eos}
a_\mu =
- \Tr \biggl[\left(\frac{\partial H }{\partial A_\mu}\right) \frac{\delta(E-H)}{\omega}\biggr]
&=
-\frac{1}{\omega}\Tr \biggl[-\frac{\partial }{\partial A_\mu} \Theta(E-H)\biggr]
\\&=
\frac{1}{\omega}\frac{\partial }{\partial A_\mu}\Tr \biggl[\Theta(E-H)\biggr]
=
\TG \left(\frac{\partial \SG}{\partial A_\mu}\right).
\end{split}
\end{equation}
This proves that only the pair $(\rho, \SG)$ constitutes a consistent thermostatistical model based on the MC density $\rho$. As a corollary, the Boltzmann $\SB$ is not a thermodynamic entropy of the MC ensemble.

\par
In a similar way,  one can show by a straightforward calculation that, for standard  classical Hamiltonian systems, only the Gibbs temperature $\TG$ satisfies the mathematically rigorous\footnote{The direct proof  of \eqref{e:gibbs_equi} requires mild assumptions such as confined trajectories and a finite groundstate energy. The key steps are very similar to those in~\eqref{e:gibbs_eos}, i.e., one merely needs to exploit the chain rule relation \mbox{$\partial \Theta(E-H)/\partial \lambda=-(\partial H/\partial \lambda)\delta(E-H)$}, which holds for any variable $\lambda$ appearing in the Hamiltonian $H$.} equipartition theorem~\citep{Khinchin}  
\begin{equation}
\label{e:gibbs_equi}
\left\langle \xi_i \frac{\partial H }{\partial \xi_j}\right\rangle
\equiv
\Tr \biggl[\left( \xi_i \frac{\partial H }{\partial \xi_j}\right) \rho\biggr]
=
\kB \TG\, \delta_{ij}
\end{equation}
for all canonical coordinates $\xi=(\xi_1,\ldots)$. Equation~\eqref{e:gibbs_equi} is essentially a phase-space version of  Stokes' theorem, relating a surface (flux) integral on the energy shell to the enclosed phase space volume.
\par

\subsection{Basic examples}
\label{s:MCE_examples}

\paragraph{Ideal gas.}
The differences between $\SB$ and $\SG$ are negligible for most macroscopic systems with monotonic DoS~$\omega$, but can be significant for small systems. This can already be seen for a classical ideal gas in $d$-space dimensions, where~\citep{2006DuHi}
\begin{equation}
\label{e:ideal_gas}
\Omega(E,V)=\alpha E^{dN/2}V^N,
\qquad
\alpha=\frac{(2\pi m)^{dN/2}}{N!h^d\Gamma(dN/2+1)},
\end{equation}
for $N$ identical particles of mass $m$ and Planck constant $h$. From this, one finds that only the Gibbs temperature yields exact equipartition
\begin{align}
\label{e:T_B_equi}
E &= \left( \frac{dN}{2}-1\right)\kB\TB,
\\
\label{e:T_G_equi}
E &= \frac{dN}{2}\kB\TG.
\end{align} 
Note that Eq.~\eqref{e:T_B_equi} yields a paradoxical results for $dN=1$, where it predicts a negative temperature $\TB<0$ and heat capacity, and also for $dN=2$, where the temperature $\TB$ must be infinite. This is a manifestation of the fact that the Boltzmann entropy $\SB$ is not an exact thermodynamic entropy. By contrast, the Gibbs entropy $\SG$ produces the reasonable result~\eqref{e:T_G_equi}, which is a special case of the more general theorem~\eqref{e:gibbs_equi}.

\paragraph{}
That $\SG$ also is the more appropriate choice for isolated quantum systems, as relevant to the interpretation of the experiments Purcell and Pound~\citep{1951PurcellPound} and Braun {\it et~al.}~\citep{2013Braun}, can be readily illustrated by two basic examples:

\paragraph{Quantum oscillator.}
 For a simple harmonic oscillator with spectrum 
\begin{equation}
E_n=\hbar\nu \left(n+\frac{1}{2}\right), 
\qquad 
n=0,1,\ldots,\infty
\end{equation}
we find by inversion and analytic interpolation $\Omega=1+n=1/2+E/(\hbar\nu) $  and, hence,  from the Gibbs entropy $\SG=\kB\ln \Omega$ the caloric equation of state 
\begin{equation}
\kB\TG=\frac{\hbar\nu}{2}+E,
\end{equation} 
which when combined with the quantum virial theorem yields an equipartition-type statement\footnote{In contrast to classical Hamiltonian systems, equipartition is not a generic feature of quantum systems, but a consistent thermodynamic formalism should be able to confirm its presence or absence also for quantum systems.} for this particular example.  Furthermore,  $T=\TG$  gives a sensible predictions for the heat capacity,
\begin{equation}
C=\left(\frac{\partial T}{\partial E}\right)^{-1} =\kB,
\end{equation} 
accounting for the fact that even a single oscillator can serve as minimal quantum heat reservoir. More precisely, the energy of a quantum oscillator can be changed by performing work through a variation of its frequency $\nu$, or by injecting or removing energy quanta, corresponding to heat transfer in the thermodynamic picture. The Gibbs entropy $\SG$ reflects these facts correctly.

By contrast, the Boltzmann entropy $\SB= \kB \ln(\epsilon\omega)$ with $\omega=(\hbar\nu)^{-1}$ assigns the same constant entropy\footnote{This value can be normalized to zero by the particular choice $\epsilon=\hbar\nu$. Generally,  the result $\SB=0$ for non-degenerate oscillator states indicates two things: (i) The Boltzmann entropy $\SB$ is a  (particular form of) information entropy, and (ii) it is not a thermodynamic entropy; see  Sec.~\ref{s:myths} below for a more detailed discussion of this issue.} to all energy states, yielding the nonsensical result $\TB=\infty$ for all energy eigenvalues~$E_n$ and making it impossible to compute the heat capacity of the oscillator. That the Boltzmann entropy  $\SB$ fails for this basic example should raise serious doubts about its applicability to more complex quantum systems.

\paragraph{Quantum particle in a box.}
The fact that $\SB$ violates fundamental thermodynamic relations, such as Eq.~\eqref{e:consistency_V}, not only for classical but also for quantum systems can be further illustrated by another elementary example. Considering a quantum particle in a one-dimensional infinite square-well of length $L$, the spectral formula 
\begin{equation}
\label{e:ideal_gas_1}
E_n=a n^2/L^2, 
\qquad 
n=1,2,\ldots,\infty
\end{equation}
implies $\Omega=n=L\sqrt{E/a}$. In this case, the Gibbs entropy $\SG=\kB \ln\Omega$  gives
\begin{equation}
\kB \TG = 2E,\qquad
p_G \equiv \TG\left(\frac{\partial \SG}{\partial L}\right)
= \frac{2E}{L},
\end{equation}
as well as the heat capacity $C=\kB/2$, in agreement with physical intuition. 
In particular, the pressure equation consistent with condition~\eqref{e:consistency_V},
as can bee seen by differentiating~\eqref{e:ideal_gas_1} with respect to $L$,
\begin{equation}
p\equiv-\frac{\partial E}{\partial L}=\frac{2E}{L} =\pG.
\end{equation}
\par
By contrast, we find from $\SB= \kB \ln(\epsilon\omega)$ with  $\omega=L/(2\sqrt{E a})$ for the Boltzmann temperature\footnote{One sometimes encounters the \textit{ad hoc} convention that, because the spectrum \eqref{e:ideal_gas_1} is non-degenerate, the \lq{}thermodynamic\rq{} entropy should be zero, $\SB=0$, for all states. However, this postulate leads to several other inconsistencies, which are discussed in more detail in Sec.~\ref{s:myths}. Focussing on the example at hand for the moment, let us just note that $\SB=0$ would again imply the nonsensical result $\TB=\infty$, misrepresenting the physical fact that also a single degree of freedom in a box-like confinement can store heat in finite amounts.} 
\begin{equation}
\kB \TB = -2E <0.
\end{equation}
While this result in itself seems questionable\footnote{Unless one believes that a quantum particle in a one-dimensional box is a Dark Energy candidate.}, it also implies a violation of Eq.~\eqref{e:consistency_V}, 
since
\begin{equation}
\pB \equiv \TB\left(\frac{\partial \SB}{\partial L}\right)=-\frac{2E}{L} \ne p.
\end{equation}
This contradiction corroborates that $\SB$ cannot be the correct entropy for quantum systems.  

\paragraph{}
We hope that the arguments and examples presented thus far suffice to convince the reader that 
the Boltzmann entropy $\SB$ is not a consistent thermodynamic entropy, neither for classical 
nor for quantum systems, whereas the Gibbs entropy $\SG$ provides a consistent thermodynamic formalism in the low energy limit (small quantum systems) and in the high-energy limit (classical systems).
\par
Unfortunately, the Boltzmann entropy has become so widely accepted nowadays that, even when its application to exotic new states leads to spectacular claims, these are hardly ever questioned 
anymore. In Sec.~\ref{s:spin_example}, we demonstrate  by means of a slightly more elaborate 
example, how naive usage of $\SB$ can lead to \lq negative temperatures that are hotter than the hottest  positive temperatures\rq\footnote{Spurious arguments, often encountered in attempts to proclaim $\SB$ as superior to the Gibbs proposal $\SG$, will be addressed in Sec~\ref{s:myths}.}.

\subsection{Measuring $\TB$  \textit{vs.} $\TG$}
To conclude this section, let us still clarify  why the method employed, for example, by Braun {\it et al.}~\citep{2013Braun} measures $\TB$ and not the thermodynamic temperature $T=\TG$.
 We restrict ourselves to sketching the main idea as the technical details of the derivation can be found in most modern textbooks on statistical mechanics~\citep{Becker,Huang}.

We recall that Braun {\it et al.}~\citep{2013Braun} estimate an effective \lq{}temperature\rq{} by fitting an exponential Bose-Einstein function to their experimentally obtained one-particle distributions. 
Let us assume their system contains $N\gg 1$ particles and denote the corresponding Hamiltonian by $H_N$ and the DoS by $\omega_N$. Then, the formally exact MC one-particle density operator is given by 
\begin{equation}
\label{e:rho_1}
\rho_1
= \Tr_{N-1}[\rho_N]
=\frac{\Tr_{N-1}[\delta(E-H_N)]}{\omega_N}.
\end{equation}
To obtain an exponential (canonical)  fitting formula, as used in the experiments, one first has to rewrite
$\rho_1$ in the equivalent form
\begin{equation}
\rho_1=\exp[\ln \rho_1].
\end{equation}
Then,  applying a standard steepest descent approximation~\citep{Becker,Huang} to the logarithm and assuming discrete one-particle levels $E_\ell$, one finds for the relative occupancy $p_\ell$ of one-particle level $E_\ell$  the canonical form\footnote{This becomes obvious by writing~\eqref{e:rho_1} for a given one-particle energy $E_\ell$ as $p_\ell=\omega_{N-1}(E-E_\ell)/\omega_N(E)=\exp[\ln\omega_{N-1}(E-E_\ell)]/\omega_{N}(E)$ and expanding for $E_\ell\ll E$, which gives $p_\ell \propto \exp[-E_\ell/(\kB T_{B,N-1})]$ where $\kB T_{B,N-1}\equiv\omega_{N-1}(E)/\omega_{N-1}'(E)$ in agreement with Eq.~\eqref{e:T_B}. That is, $\TB$ in~\eqref{e:canonical} is actually the Boltzmann temperature of the $(N-1)$-particle system.}
\begin{equation}
\label{e:canonical}
p_\ell \simeq \frac{e^{-E_\ell/(\kB \TB)}}{Z},
\qquad Z=\sum_{\ell} e^{-E_\ell/(\kB \TB)}.
\end{equation}
The key observation here is that this exponential approximation features~$\TB$  and not the absolute thermodynamic Gibbs temperature~$T=\TG$. Hence, by fitting the one-particle distribution, Braun {\it et~al.}~\citep{2013Braun} determined the Boltzmann temperature $\TB$, which can be negative, whereas the thermodynamic Gibbs temperature $T=\TG$ is always non-negative. From the above definitions, it is straightforward to show that, generally,  
\begin{equation}\label{e:rel_vs_abs}
\TB=\frac{\TG}{1- \kB/C},
\end{equation}
where  $C=(\partial \TG/\partial E)^{-1}$ is the heat capacity. Evidently, differences between $\TG$ and $\TB$ become relevant only if $|C|$ is close to or smaller than $\kB$; in particular,  $\TB$ is negative if \mbox{$0<C<\kB$}. From a practical perspective,  Eq.~\eqref{e:rel_vs_abs} is useful as it allows to reconstruct the non-negative absolute temperature~$T=\TG$ from measurements of~$\TB$ and~$C$,
but $\TG$ can, of course, also be directly measured. 
\par
For example, the equipartition theorem \eqref{e:gibbs_equi} for classical systems implies that an isolated ideal gas thermometer shows,  strictly speaking, the Gibbs temperature $\TG$ and not\footnote{Of course, for most macroscopic systems, $\TG$ and $\TB$ are practically indistinguishable, see~\eqref{e:T_B_equi} and \eqref{e:T_G_equi}.} the Boltzmann temperature~$\TB$. Furthermore, when brought into (weak) thermal contact with an otherwise isolated system, a gas thermometer  shows the Gibbs temperature of the compound system, not $\TB$. 
For completeness, we describe in App.~\ref{quantum_thermometer} a simple protocol for how one can directly measure $\TG$ for quantum systems in practice. 
\par
In summary,  the Gibbs entropy provides not only the consistent thermostatistical description of isolated systems but also a sound practical basis for classical and quantum thermometers.
\pagebreak

\section{Generic example with bounded spectrum}
\label{s:spin_example}
\begin{figure*}
\includegraphics[width=15.5cm]{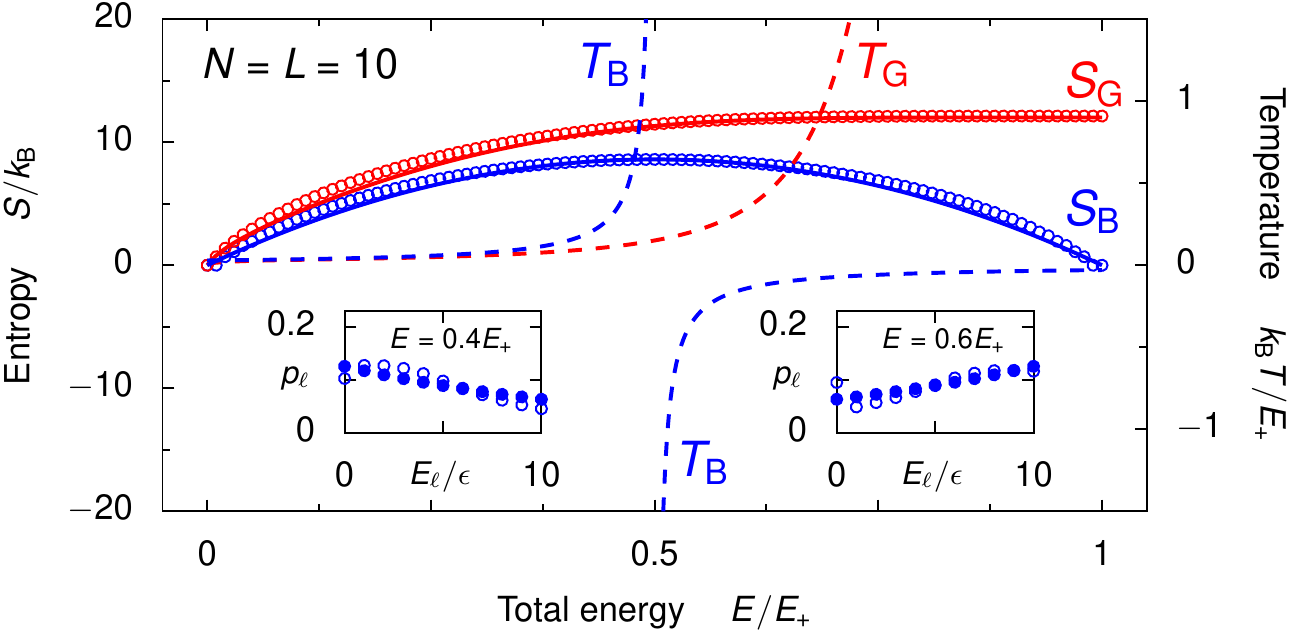}
\caption{Non-negativity of the absolute temperature in quantum systems with bounded spectrum.
Thermodynamic functions for $N$ weakly coupled bosonic oscillators with $(L+1)$ single particle levels $E_\ell=\ell\epsilon$, $\ell =0,\ldots, L$ are shown for $N=L=10$, corresponding to $184756$ states in the energy band $[E_-,E_+]=[0,LN\epsilon]$. Open circles show exact numerical data; lines represent analytical results based on the Gaussian approximation of the DoS $\omega$. The thermodynamic Gibbs entropy $S=\SG=\kB\ln \Omega$ (red solid) grows monotonically with the total energy $E$, whereas the Boltzmann (or surface) entropy $\SB=\kB\ln (\epsilon \omega)$ (blue solid) does not. Accordingly, the absolute temperature $T=\TG$ (red dashed) remains positive, whereas the Boltzmann temperature $\TB$ (blue dashed), as measured by Braun {\it et~al.}~\citep{2013Braun}, exhibits a singularity at $E_*=\epsilon NL/2$.  
Note that, although $\TG$ increases rapidly for $E>E_*/2$, it remains finite since $\omega(E)>0$ on $[0,E_+]$. Insets: Exact relative occupancies $p_\ell$ (open circles) of one-particle energy levels are shown for two different values of the total energy. They agree qualitatively with those in Figs.~1A  and 3 of Ref.~\citep{2013Braun}, and can be approximately reproduced by an exponential distribution (filled circles) with parameter $\TB$, see Eq.~\eqref{e:canonical}. Quantitative deviations result from the limited sample size $(N,L)$ and the use of the Gaussian approximation for $\TB$ in our model calculations.
\label{fig:thermodynamic_functions_for_weakly_coupled_bosonic_oscillators}
}
\end{figure*}

That the difference between $\TG$ and $\TB$ is practically negligible for conventional macroscopic systems~\citep{Becker,Huang} may explain why  they are rarely distinguished in most modern textbooks apart from a few exceptions~\citep{Khinchin,Becker}. However, for quantum systems with  bounded energy spectrum,  $\SG$ and $\SB$ are generally very different (Fig.~\ref{fig:thermodynamic_functions_for_weakly_coupled_bosonic_oscillators}), and a careful distinction between $\TG$ and $\TB$ becomes necessary.

To demonstrate this, we consider a generic quantum model that formalizes the example presented by Braun {\it et~al.}~\citep{2013Braun}  in Fig.~\ref{fig:thermodynamic_functions_for_weakly_coupled_bosonic_oscillators}A of their paper\footnote{When interpreted in terms of spins, this model  applies also to the experiments of~Purcell and Pound~\citep{1951PurcellPound}.}. The model consists of $N$ weakly interacting bosonic oscillators with Hamiltonian 
\begin{equation}
{H}_N\simeq \sum_{n=1}^N {h}_n,
\end{equation}
such that each oscillator can occupy non-degenerate single-particle energy levels $E_{\ell_n}=\epsilon \ell_n$ with spacing $\epsilon$  and $\ell_n =0,1\ldots, L$. Assuming indistinguishable bosons, permissible $N$-particle states can be labelled by $\Lambda=(\ell_1,\ldots,\ell_N)$, where $0\le\ell_1\le \ell_2\ldots\le \ell_N\le L$, and the associated energy eigenvalues  $E_\Lambda=\epsilon(\ell_1+\ldots+\ell_N)$ are bounded by $0\le E_\Lambda \le E_+=\epsilon LN$. The DoS 
\begin{equation}
\omega_N(E)=\Tr_{N}[\delta(E-H_N)]
\end{equation}
counts the degeneracy of the eigenvalues $E$ and equals the number of integer partitions~\citep{Stanley} of $z=E/\epsilon$ into $N$ addends $\ell_n\le L$. For $N,L\gg 1$, the DoS  can be approximated by a continuous Gaussian, 
\begin{equation}
\omega(E)=\omega_* \exp[-(E- E_*)^2/\sigma^2],
\end{equation}
and the degeneracy attains its maximum $\omega_*$ at the center  $E_*=E_+/2$ of the energy band (Fig.~\ref{fig:thermodynamic_functions_for_weakly_coupled_bosonic_oscillators}).   The integrated DoS is then obtained as
\begin{equation}
\begin{split}
\Omega(E) &=
\Tr_{N}[\Theta(E-H_N)] 
\\&\simeq
 1+\int _0^E \omega(E') dE'
\\&=
1+\frac{\omega_*\sqrt{\pi}\sigma}{2}
\left[\mathrm{erf}\biggl( \frac{E-E_*}{\sigma}\biggr)+\mathrm{erf}\biggl( \frac{E_*}{\sigma}\biggr)\right] ,
\end{split}
\end{equation}
where the parameters $\sigma$ and $\omega_*$ are determined by  the boundary condition $\omega(0)=1/\epsilon$ and the total number~\citep{Stanley} of possible $N$-particle states $\Omega(E_+)=(N+L)!/(N!L!)$.  For this  model, the Gaussian approximation gives 
\begin{equation}
\kB \TB= \frac{\sigma^2}{E_+ -2E}
\end{equation}
which diverges and changes sign as $E$ crosses $E_*=E_+/2$, whereas the absolute temperature $T=\TG(E)$ grows monotonically but remains finite (Fig.~\ref{fig:thermodynamic_functions_for_weakly_coupled_bosonic_oscillators}). In a quantum system with bounded spectrum as illustrated in Fig. 1, the heat capacity $C$ decreases rapidly towards $\kB$ as the energy approaches  $E_*=E_+/2$, and $C$ does not to scale homogeneously with system size anymore as $E\to E_+$ due to combinatorial constraints on the number of available states.  Note that, for the same reason, the weak coupling assumptions underlying derivations of canonical distributions, as in Eq.~\eqref{e:canonical},  become invalid  as one approaches $E_+$.
\par
In summary, for systems with bounded spectrum similar to the one in Fig.~\ref{fig:thermodynamic_functions_for_weakly_coupled_bosonic_oscillators}, the effective Boltzmann temperature $\TB$ differs
not only quantitatively but also qualitatively from the true thermodynamic temperature $\TG>0$.

\section{Myths and facts}
\label{s:myths}
The considerations in the previous sections demonstrate that the Boltzmann entropy, when interpreted as a  thermodynamic entropy,  leads to a number of inconsistencies, whereas the Gibbs entropy respects the thermodynamic relations~\eqref{e:fundamental} and also gives reasonable  
results for both small quantum  and arbitrary classical Hamiltonian systems. Notwithstanding, in various discussions over the last decade, we have met a number of recurrent arguments opposing the Gibbs entropy as being conceptually inferior to the Boltzmann entropy. None of those objections, however, seems capable of withstanding careful inspection. We therefore thought it might be helpful to list, and address explicitly, the most frequently encountered arguments that may seem plausible at first but turn out to be unsubstantiated.

\paragraph{\textit{The Gibbs entropy violates the second law $dS\ge 0\;$ for closed systems, whereas  the Boltzmann entropy does not.}} This statement is incorrect, simply because for closed systems with fixed control parameters (i.e., constant energy, volume, etc.) both Gibbs and Boltzmann entropy are constant.
This general fact, which follows trivially from the definitions of $\SG$ and $\SB$, 
is directly illustrated by  the classical ideal gas example discussed above.

\paragraph{\textit{Thermodynamic entropy must be equal to information entropy, and this is true only for the Boltzmann entropy.}}
This argument can be discarded for several reasons. Clearly, entropic information measures themselves are a matter of convention~\citep{1960Renyi}, and there exists a large number of different entropies (Shannon, Renyi, Kuhlback entropies, etc.), each  having their own virtues and drawbacks as measures of information~\citep{1978Wehrl}. However,  only few of those entropies, when combined with an appropriate probability distribution, define ensembles that obey the fundamental thermodynamic relations~\eqref{e:fundamental}. It so happens that the entropy of the canonical ensemble\footnote{More precisely, one should say that the exponential  (canonical) Boltzmann distribution combined with Shannon's entropy forms a consistent thermostatistical model of thermodynamics, as it is possible to identify meaningful expectation values that satisfy the fundamental differential relations~\eqref{e:fundamental} of thermodynamics;  see Ref.~\citep{2007Campisi} for a discussion of other examples.} coincides with Shannon's popular information measure. But the canonical ensemble (infinite bath)  and the more fundamental MC ensemble (no bath) correspond to completely different physical situations~\citep{Becker} and, accordingly, the MC entropy is, in general, not equivalent to Shannon's information entropy\footnote{Except in those limit cases where MC and canonical ensembles become equivalent.}. Just by considering classical Hamiltonian systems, one can easily verify that neither the Boltzmann nor the Gibbs entropy belong to the class of Shannon entropies\footnote{This does not mean that these two different entropies cannot be viewed as measures of information. Both Gibbs and Boltzmann entropy encode valuable physical information about the underlying energy spectra, but only one of them agrees with thermodynamics.}. Moreover, there is no absolute need for identifying thermodynamic and information-theoretic entropies\footnote{For a clear view on this topic, see  the discussion on p.142 in Khinchin's textbook~\citep{Khinchin}.}. Although it may seem desirable to unify information theory and thermodynamic concepts for formal or aesthetic reasons, some reservation is in order~\citep{Khinchin} when such attempts cause mathematical inconsistencies and fail to produce reasonable results in the simplest analytically tractable cases.


\paragraph{\textit{Non-degenerate states must have zero thermodynamic entropy, and this is true only for the Boltzmann entropy.}}
This argument again traces back to confusing thermodynamic and information entropies~\citep{Khinchin}. Physical systems that possess non-degenerate spectra can be used to store energy, and
one can perform work on them by changing their parameters. It seems reasonable to  demand that a well-defined thermodynamic  formalism is able to account for these facts. Hence, entropy definitions that are insensitive to the full energetic structure of the spectrum  by only counting degeneracies of individual levels are not particularly promising candidates for 
capturing thermodynamic properties.
\par
Moreover, it is not true that the Boltzmann entropy, when defined with respect to a coarse-grained DoS, as commonly assumed in applications, assigns zero entropy to non-degenerate spectra, as the DoS merely measures the total number of states in predefined energy intervals but does not explicitly reflect the degeneracies of the individual states.
\par
If, however, one were to postulate that the thermodynamic entropy of an energy level $E_n$  with degeneracy $g_n$ is exactly equal to $\kB\ln g_n$, then this would lead to other undesirable consequences: Degeneracies usually reflect symmetries that can be broken by infinitesimal parameter variations. That is, if one were to adopt $\kB\ln g_n $, then the entropy of the system could be set to zero,  for many or even all energy levels, just by a very small parameter variation that lifts the exact degeneracy\footnote{E.g., small defects in a sample that destroy exact degeneracies.}, even though actual physical properties (e.g., heat capacity, conductivity) are not likely to be that dramatically affected by minor deviations from the exact symmetry. By contrast, an integral measure such as the Gibbs entropy responds much more continuously (although not necessarily smoothly) to such infinitesimal changes.\footnote{Note that the ground-state $E_0$ is an exception since, in this case, $\SG(E_0)=\kB\ln\Omega(E_0)=\kB\ln g_0$, where $g_0$ is the ground-state degeneracy; i.e., the Gibbs entropy agrees with the experimentally confirmed residual entropy~\cite{1928Giauque,1935Pauling}, see footnote on p.8 in Ref.~\cite{2009Campisi_JPhysA}.}

\pagebreak
\paragraph{\textit{If the spectrum is invariant under $E\to-E$,  then so should be the entropy.}}
At first sight, this statement may look like a neat symmetry argument in support of the Boltzmann entropy, which indeed exhibits this property  (see example in Fig.~\ref{fig:thermodynamic_functions_for_weakly_coupled_bosonic_oscillators}). However, such an additional axiom would be in conflict with the postulates of traditional thermodynamics, which require $S$ to be a monotonic function of the energy~\citep{Callen}. On rare occasions, it can be beneficial or even necessary to remove, replace and adapt certain axioms even in a well-tested  theory,  but such radical steps need to be justified by substantial experimental evidence. The motivation for the \lq new\rq~entropy invariance postulate is merely originating from the vague idea that, for systems with a DoS as shown in Fig.~\ref{fig:thermodynamic_functions_for_weakly_coupled_bosonic_oscillators}, the maximum energy state (\lq all spins up\rq) is somehow equivalent to the lowest energy state (\lq all spins down\rq). Whilst this may be correct if one is only interested in comparing degeneracies, an experimentalist who performs thermodynamic manipulations will certainly be able to distinguish the groundstate from the highest-energy state\footnote{The groundstate can absorb photons, whereas the maximum energy state cannot.}. Since thermostatistics should be able to  connect experiment with theory, it seems reasonable to maintain that the thermodynamic entropy should reflect the absolute difference between energy-states.

\paragraph{\textit{Thermodynamic relations can only be expected to hold for large systems, so it is not a problem that the Boltzmann entropy does not work for small quantum systems.}}
Apart from the fact that the Boltzmann entropy does not obey the fundamental thermodynamic relation~\eqref{e:fundamental},  it seems unwise to build a theoretical framework on postulates that 
fail in the simplest test cases, especially, when Gibbs' original proposal~\citep{Gibbs} appears to work perfectly fine for systems of arbitrary size\footnote{A practical \lq advantage\rq~of large systems is that thermodynamic quantities typically become \lq sharp\rq~\citep{Becker} when considering a suitably defined thermodynamic limit, whereas for small systems fluctuations around the mean values are relevant. However, this does not mean that thermostatistics itself must become invalid for small systems. In fact, the Gibbs formalism~\citep{Gibbs} works perfectly fine even for small MC systems~\citep{2006DuHi,2007Campisi},  and a logically correct statement would be: The Boltzmann entropy produces reasonable results for a number of large systems because it happens to approach the thermodynamically consistent Gibbs entropy in those (limit) cases.}. To use two slightly provocative analogies: It does not seem advisable to replace the Schr\"odinger equation by a theory that fails to reproduce the hydrogen spectrum but claims to predict more accurately the spectral properties of larger quantum systems. Nor would it seem a good idea to trust a numerical algorithm that produces exciting results for large systems but fails to produce sensible results for one- or two-particle test scenarios. If one applies similar standards to the axiomatic foundations of thermostatistics, then the Boltzmann entropy should be replaced by the Gibbs entropy~\eqref{e:S_G}, implying that negative absolute temperatures cannot be achieved.

\paragraph{\textit{The Gibbs entropy is probably correct for small quantum systems and classical systems, but one should use the Boltzmann entropy for intermediate quantum systems.}}
To assume that a theoretical framework that is known to be inconsistent in the low-energy limit of small quantum systems as well as in the high-energy limit of classical systems, may be preferable in some intermediate regime  seems adventurous at best.

\paragraph{}
We hope that the discussion in this section, although presented in an unusual form,  is helpful for the objective evaluation of Gibbs and Boltzmann entropy\footnote{One could add two more \lq arguments\rq~to the above list: (i) \lq The Boltzmann entropy is prevalent in modern textbooks and has been more frequently used for more than 50 years and, therefore, must be correct\rq -- we do not think such reasoning is constructive. (ii) \lq The Gibbs entropy gives incorrect results for simple systems such as the ideal gas, etc.\rq -- this can be easily disproven with Eq.~\eqref{e:ideal_gas} and similarly elementary calculations for other systems.}.  It should be emphasized, however, that no false or correct argument against the Gibbs entropy can cure the thermodynamic incompatibility of the Boltzmann entropy.

\section{Carnot efficiencies $>1\,$?}
\label{s:carnot}

For completeness, we still comment briefly on speculations~\citep{1956Ramsay,2010Rapp,2013Braun} that population-inverted systems can provide Carnot machines with efficiency $>1$.
To evaluate such statements, let us recall that a Carnot cycle, by definition, consists of four successive steps: (I) isothermal expansion; (II) isentropic  expansion; (III) isothermal compression; (IV) isentropic compression. Steps I and III require a hot  and cold bath with temperatures $\THot$ and $\TCold$, respectively, and the isentropic steps II and IV can be thought of as place-holders for other more general work-like parameter variations (changes of external magnetic fields, etc.). The associated Carnot efficiency, defined by 
\begin{equation}\label{e:eta}
\eta=1-\frac{\TCold}{\THot}, 
\end{equation}
owes its popularity to the fact that it presents an upper bound for other 
heat engines. To realize values $\eta>1$, one requires either $\TCold$ or $\THot$ to be negative.
At least formally,  this appears to be achievable by considering a spectrum as in Fig.~\ref{fig:thermodynamic_functions_for_weakly_coupled_bosonic_oscillators}
and naively inserting positive and negative Boltzmann temperature values into  Eq.~\eqref{e:eta}.
\par
We think that  speculations~\citep{1956Ramsay,2010Rapp,2013Braun} of this type are misleading for a number of reasons. First, the Boltzmann temperature  $\TB$ is not a consistent thermodynamic temperature, and, if at all, one should use the Gibbs temperature $\TG$ in Eq.~\eqref{e:eta} instead.
Second, in order to change back and forth between population-inverted states with $\TB<0$ and non-inverted states with $\TB>0$, work must be performed  in a non-adiabatic~\cite{1975Tremblay} manner (e.g., by rapidly switching a magnetic field), regardless of whether one considers Boltzmann or Gibbs entropy. That is, the resulting process is not of the Carnot-type anymore, requiring a carefully performed energy balance calculation~\citep{1977Landsberg_JPhysA}. In particular,  such an analysis has to account for the peculiar fact that, when the heat engine is capable of undergoing population-inversion,  then both hot and cold bath may inject heat into the system. Properly defined efficiencies of thermodynamic cycles that involve systems with lower and upper energy bounds are, in general, not just simple functions of $\TG$ or $\TB$. For these reasons, the naive application of  Eq.~\eqref{e:eta} can be severely misleading in those cases.

\section{Conclusions} 
Groundbreaking experiments like those by Purcell and Pound~\citep{1951PurcellPound} and  Braun {\it et~al.}~\citep{2013Braun} are essential for verifying the conceptual foundations of thermodynamics and thermostatistics. Such studies  disclose previously unexplored regimes, thereby enabling us to test and, where necessary, expand theoretical concepts that will allow us to make predictions and are essential for the development of new technologies. However, the correct interpretation of data and the consistent formulation of heat and work exchange~\citep{2009Campisi_PRL}  under extreme physical conditions (e.g., at ultracold or ultrahot temperatures, or on atomic or astronomical scales) require special care~\citep{2009DuHaHi} when it comes to applying the definitions and conventions that constitute a specific theoretical framework. When interpreted within a consistent thermostatistical theory, as developed by Gibbs~\citep{Gibbs} more than a century ago,  the pioneering experiments of both  Purcell and Pound~\citep{1951PurcellPound} and Braun {\it et~al.}~\citep{2013Braun}  suggest that the answer to the  question \lq{}Negative absolute temperatures?\rq{} should remain: \lq{}Not in thermodynamics.\rq{}
\par
To end on a conciliatory note, we do not question that  alternative temperature concepts (e.g., effective spin temperatures) can be very useful if their terms-of-use have been well-defined, and agreed upon, in a specific context -- they should however be carefully distinguished from the absolute  thermodynamic temperature~$T$, especially when misidentification causes unnecessary confusion about the validity of well-established thermostatistical axioms and theorems~\citep{Khinchin}, such as the non-negativity of~$T$ or the efficiencies of heat engines.

\paragraph{Acknowledgements.} We thank I. Bloch and U. Schneider for discussions.  
We are grateful to M. Campisi for pointing out Eq.~\eqref{e:rel_vs_abs}, and to P. H\"anggi and 
R. E. Goldstein for  numerous helpful comments.

\bibliography{NegTemp}
\bibliographystyle{unsrt}



\appendix
\section{Minimal quantum thermometer}
\label{quantum_thermometer}

A simple, if not the simplest, quantum thermometer for measuring the thermodynamic Gibbs temperature~\mbox{$T=\TG$} can be realized with a heavy atom in a 1D harmonic trap.  
The measurement protocol is as follows: Before coupling thermometer and system, we prepare the isolated system in a state with well-defined energy
\begin{equation}
E=E^S
\end{equation}
and the thermometer oscillator (frequency $\nu$, eigenstates $E^T=\hbar\nu m$) in the groundstate \mbox{$E^T=0$}. We next establish contact between thermometer and system. After the coupling, redistribution of energy via weak interaction takes place but the total energy remains conserved
\begin{equation}
E^{T+S}=\hbar\nu m +E^{\prime S} =E,
\end{equation}
with  $E^{\prime S}$ denoting the energy remaining in the system. After separating the thermometer from the system, the oscillator will be in one of the states $m=0,\ldots, M$ where $M=[E/(\hbar\nu)]$, with $[x]$ denoting the integer part of $x$. If the experimental setup realizes a microcanonical ensemble, the probability of finding a specific oscillator value $m$ after decoupling is given by 
\begin{equation}
P[m|E]=\frac{g(E-\hbar\nu m)}{\Omega(E)},
\end{equation}
where $g(E^{\prime S})$ is the degeneracy of the level $E^{\prime S}$ of the isolated system, and 
\begin{equation}
\Omega(E)=\sum_{E^{\prime S}\le E}g(E^{\prime S}).
\end{equation} 
Assuming that the levels lie sufficiently dense, we can approximate the discrete probabilities $P[\mu|E]$ by the probability density
\begin{equation}
p(\mu|E)=\frac{\omega(E-\mu)}{\Omega(E)},
\end{equation}
where $\mu=\hbar\nu m$ is the oscillator energy.  This distribution can be obtained by repeating the experiment many times, and a simple exact estimator for the (inverse) absolute temperature $T> 0$ is [compare Eq.~\eqref{e:T_G} above]
\begin{equation}\label{e:thermometer}
\frac{1}{\kB T}=\frac{\omega(E)}{\Omega(E)}=p(0|E).
\end{equation}
In practice, we would measure $p(\mu|E)$ for $\mu>0$ and extrapolate to $\mu=0$. The thermometer equation~\eqref{e:thermometer} is applicable to systems with and without population inversion; the precision of this minimal thermometer is set by the oscillator frequency $\nu$ and the number of measurements.

\end{document}